\documentclass[aps,prb,twocolumn,superscriptaddress,
showpacs,floatfix]{revtex4-1}
\usepackage[caption=false]{subfig}
\usepackage{graphicx}
\usepackage{dcolumn}
\usepackage{bm}
\usepackage{calc} 
\usepackage{amsfonts} 
\usepackage{amssymb} 
\usepackage{amsmath}
\usepackage{subfig} 
\usepackage{natbib}
\usepackage{array,color}
\usepackage{amsbsy}

\begin{document}

\title{The Seebeck coefficient in correlated  low dimensional organic metals}

\author{M. Shahbazi} 
\email{maryam.shahbazi@usherbrooke.ca}
\author{C. Bourbonnais}
\email{claude.bourbonnais@usherbrooke.ca }
\affiliation{%
Regroupement Qu\'ebecois sur les Mat\'eriaux de Pointe, D\'epartement de physique, Universit\'e de Sherbrooke, Sherbrooke, Qu\'ebec, Canada, J1K-2R1}%

\date{\today}

\begin{abstract}
We study the influence of inelastic electron-electron scattering on the temperature variation of the Seebeck coefficient in the normal phase of quasi-one-dimensional organic superconductors. The theory is based on the numerical solution  of the semi-classical Boltzmann equation for which the collision integral equation is solved with the aid of the electronic umklapp scattering vertex  calculated by the renormalization group method. We show that the one-loop renormalization group  flow of momentum and temperature dependent umklapp scattering, in the presence of nesting alterations of the Fermi surface, introduce electron-hole asymmetry in the energy dependence of the anisotropic scattering time. This is responsible for the enhancement of the Seebeck coefficient with respect to the band $T$-linear prediction and even its sign reversal around the quantum critical point of the phase diagram, namely where the interplay between antiferromagnetism and superconductivity along with the strength of spin fluctuations are the strongest. Comparison of the results with available data on low dimensional organic superconductors is presented and critically discussed.

\end{abstract}

\maketitle
\section{Introduction}
In the past few years we have seen  expanding interest in the Seebeck coefficient  as a sensitive probe of fluctuations   encased in the quantum critical behaviour of correlated electrons. This has been exemplified both  experimentally  and theoretically  for quantum critical points   in  heavy fermions\cite{Hartmann10,Pfau12,Ren16,Paul01,Pepin10},   pnictides \cite{Arsenijevic13}, and  for hole \cite{Laliberte11,Badoux16,Buhmann13,Arsenault13}, and electron-doped\cite{Li07,Gollnik98} cuprates. In organic superconductors like the  Bechgaard salts (TMTSF)$_2X$  series,  also known to exhibit quantum criticality, the  measurements of the Seebeck coefficient  have been the subject of numerous reports  following their discovery \cite{Bechgaard80,Mortensen82,Chai09,Choi02,Sun08,Chaikin83,Gubser82}, and this, until very recently\cite{Machida16}. However,   these works have found very little theoretical echo  as to the possible part played by
quantum fluctuations in the thermoelectric response seen in these correlated quasi-one dimensional  (quasi-1D) metals.  This topic that has  remained  essentially unexplored so far\cite{Behnia15},   will  be  the main focuss  of the present work.

The quantum critical behaviour of the Bechgaard salts is known to result from the juncture of  a declining spin-density-wave (SDW) state  with the onset  of a superconducting (SC) dome  under pressure\cite{Jerome82,Bourbon08,Taillefer10,Brown15,Jerome16}. The signatures of quantum criticality have been chiefly revealed by the observation of linear-$T$ resistivity\cite{DoironLeyraud09}, whose strength scales with the  distance from the quantum critical point (QCP) along the pressure axis. Among other fingerprints of quantum criticality, linear resistivity  was also  found to scale with  the amplitude  of SDW fluctuations seen by NMR  and  with the size of the critical temperature $T_c$ for superconductivity\cite{Brown08,Creuzet87b,Brown15,Kimura11}.  

The contributions of the renormalization group (RG) approach to the understanding of these quantum critical features  have been the purpose of several works in the past\cite{Duprat01,Nickel06,Bourbon09}. In the framework of the quasi-1D electron gas model  for instance, that is how  the characteristic  sequence of instabilities lines and the  scaling of spin fluctuations with the size of  $T_c$ could   be fairly well simulated when  the antinesting amplitude of the quasi-1D electron band structure is used  as a tuning parameter for the QCP\cite{Bourbon09,Sedeki12,Fuseya12}.    

More recently, the   RG calculations for the   umklapp vertex was shown to serve as an input to  the linearized Boltzmann theory  of  electrical transport \cite{Shahbazi15}. From this combination of techniques, first proposed  by Buhmann {\it et al.}\cite{Buhmann13}  in the context of the 2D  Hubbard model for the cuprates, the metallic resistivity across the QCP could be calculated along   the  pressure - antinesting - axis and the results congruently compared with experiments in the Bechgaard salts\cite{DoironLeyraud09}.

In this work we further exploit  the RG-Boltzmann approach and derive the Seebeck coefficient for correlated quasi-1D metals. The numerical integration of the linearized Boltzmann equation when  fed in  by the RG  umklapp vertex function, allows a microscopic determination of the energy variation of the anisotropic electron-electron scattering time across the Fermi surface.  This variation is mostly influenced by SDW fluctuations and is anisotropic on the Fermi surface. It introduces deviations  with respect to the Seebeck coefficient obtained in the band limit,  which is  positive and linear in temperature  for hole carriers in materials like the Bechgaard salts. The  deviations  take the form of enhancements that can be not only positive,  but also negative or electron like in character.  The latter can lead to the sign reversal of the Seebeck coefficient, especially in the neighborhood of  the QCP  where the interplay between SDW and SC, together with the amplitude of the SDW fluctuations in the metallic state,  are  the strongest. These results  offer an interesting avenue  for the understanding of the sign reversal in the Seebeck coefficient  of  the  Bechgaard salts near their QCP.

The theory    is  broadened  to   systems with stronger umklapp scattering that favours a Mott  instability in the 1D portion of their metallic state, which can be approached by  the weak coupling RG from the high temperature domain. The results are confronted to the measurements of the Seebeck coefficient for prototype  members of the   sulfur based compounds, the  (TMTTF)$_2X$, known as the  Fabre salts series \cite{Mortensen83}. These compounds are characterized by  a more pronounced dimerization of the organic stacks which magnifies umklapp scattering and favours a crossover  toward a 1D Mott insulating state.  

In Sec.~II, we use the  linearized Boltzmann theory to derive the expression of the Seebeck coefficient for a quasi-1D three-quarter filled hole band of a lattice of weakly dimerized chains. In Sec.~III, the momentum-resolved renormalized umklapp vertex entering the Boltzmann equation is computed  by the RG, in the conditions of the quasi-1D electron gas model   simulating the sequence of instabilities found in     the Bechgaard salts. In Sec.~IV, we present the temperature variations of the Seebeck coefficient across the quantum critical point of the model  and examine their link with the energy profile of the inelastic scattering time.    In Sec.~V, a comparison of the results is made with available data  for   (TMTSF)$_2X$,   and on a broader basis  for the more correlated compounds (TMTTF)$_2X$. We conclude in Sec.~VI.

\section{Linearized Boltzmann theory of   the Seebeck coefficient}
We consider the semicalssical Boltzmann equation for the variation of the  quasi-particle Fermi  distribution function $f$   in the presence of collisions  and a thermal gradient $\nabla_{\bm{r}} T$. In steady-state conditions, it takes the form
\begin{align}
\label{Boltzmann}
   \left[\dfrac{\partial f(\bm{k})}{\partial t}\right]_{\text{coll}}=  e\,\boldsymbol{\cal E}\cdot {\nabla}_{\hbar \bm{k}}f -  {(E_{\bm{k}} -\mu)\over T}\nabla_{\bm{r}} T \cdot {\nabla}_{\hbar \bm{k}}f,
\end{align} 
where $E_{\bm{k}}$ is the carrier spectrum, $\mu$ the chemical potential, $e$  the electron charge and $\boldsymbol{\cal E}$ the electric field set up by the thermal gradient. The collision integral for an array of $N_P$ chains of length $L$  takes the form
\begin{align}\label{Coll}
\left[\dfrac{\partial f(\bm{k})}{\partial t}\right]_{\text{coll}}  = & -   (LN_P)^{-2} \sum \limits_{\bm{k}_2,\bm{k}_3,\bm{k}_4}{1\over 2}|\langle\bm{k},\bm{k}_2\vert g_3 \vert\bm{k}_3,\bm{k}_4\rangle \cr 
& -\langle\bm{k},\bm{k}_2\vert g_3 \vert\bm{k}_4,\bm{k}_3\rangle|^2 
   \frac{2\pi}{\hbar}  \delta_{\bm{k}+\bm{k}_2,\bm{k}_3+\bm{k}_4\pm  \bm{G}} \cr
   & \delta(E_{\bm{k}}+E_{\bm{k}_2}-E_{\bm{k}_3}- E_{\bm{k}_4}) \nonumber \\ & \times  \{{ f(\bm{k})f(\bm{k}_2)[1-f(\bm{k}_3)][1-f(\bm{k}_4)]}\cr
&- [1-f(\bm{k})][1-f(\bm{k}_2)]f(\bm{k}_3)f(\bm{k}_4)\},
\end{align}
From the Fermi Golden rule, the transition probability per unit of time is given by the matrix element $\langle\bm{k},\bm{k}_2\vert g_3 \vert\bm{k}_3,\bm{k}_4\rangle$ for longitudinal umklapp processes  where   $\bm{G}=(4k_F,0)$ is the longitudinal reciprocal lattice wave vector and $k_F$  is the 1D Fermi wavevector.

We proceed to the linearization of the Boltzmann equation by introducing\cite{Haug08}
\begin{eqnarray}
 \label{fermiNE}
f({\boldsymbol{k}})=\dfrac{1}{e^{\beta (E_{\boldsymbol{k}}-\mu)-\phi_{\boldsymbol{k}}}+1},
\end{eqnarray}
where $\phi_{\bm{k}}$ is a normalized deviation to thermal equilibrium and  ${\beta=1/k_BT}$. In the tight-binding approximation, the hole band spectrum  for   a linear  array of $N_P$ weakly coupled dimerized chains is given by
\begin{align}
\label{Ek}
E_{\bm{k}}= \sqrt{2(t^2+\delta t^2) + 2(t^2 -\delta t^2)\cos ka}\, + \epsilon_\perp(k_\perp), 
\end{align}
where $t\pm \delta t$ are the transfer integrals within and between the dimers ($\delta t>0$,$\delta t\ll t$). Here $a$ is the lattice spacing along the chains, namely the distance between  dimers. The transverse part of the hole spectrum is given by  
\begin{equation}
\label{Eperp}
\epsilon_\perp(k_\perp) =  2t_\perp \cos k_\perp d_\perp   + 2t'_\perp \cos2 k_\perp d_\perp,
\end{equation}
 where $t_\perp$ and $t'_\perp$ are the first and the second-nearest neighbour transfer integrals in the  direction perpendicular to the chains.
 
For small deviations with respect to equilibrium,  the Fermi distribution becomes 
\begin{equation}
\label{Linearfermi}
f(\bm{k}) \simeq f^0(\bm{k}) + f^0(\bm{k}) [1- f^0(\bm{k})]\phi_{\bm{k}},
\end{equation}
where $f^0(\bm{k})$ is the equilibrium   distribution at $\phi_{\bm{k}}=0$. Replacing (\ref{Linearfermi}) into Eqs (\ref{Boltzmann}) and (\ref{Coll})  leads to  the linearized Boltzmann equation
\begin{align}
\label{}
 \mathcal{L}\phi_{\bm{k}} = &\  e\beta\mathbf{\cal E}\cdot \bm{v}_{\bm{k}} - \beta^2 k_B(E_{\bm{k}}-\mu)\bm{v}_{\bm{k}}\cdot \nabla_{\bm{r}}T \cr
  \equiv &  \ \mathcal{L}\phi^{\cal E}_{\bm{k}}  -    \mathcal{L}\phi^{T}_{\bm{k}}. 
\end{align}
The collision operator $\mathcal{L}$ satisfies the integral equation
 \begin{equation}
\label{ }
 \mathcal{L}\phi^j_{\bm{k}} = \sum_{\bm{k}'}  \mathcal{L}_{\bm{k},\bm{k}'} \phi^j_{\bm{k}'}, \ \ \ \ (j= {\cal E},T).
\end{equation}
 where the kernel is given by 
 \begin{align}
 \label{CollOp}
 \mathcal{L}_{\bm{k},\bm{k}'}  = & \   (LN_P)^{-2} \sum \limits_{\bm{k}_2,\bm{k}_3,\bm{k}_4}{1\over 2}|\langle\bm{k},\bm{k}_2\vert g_3 \vert\bm{k}_3,\bm{k}_4\rangle \cr 
& -\langle\bm{k},\bm{k}_2\vert g_3 \vert\bm{k}_4,\bm{k}_3\rangle|^2 
   \frac{2\pi}{\hbar}  \delta_{\bm{k}+\bm{k}_2,\bm{k}_3+\bm{k}_4\pm  \bm{G}} \cr
   & \delta(E_{\bm{k}}+E_{\bm{k}_2}-E_{\bm{k}_3}- E_{\bm{k}_4}) \nonumber \\ & \times { f^0(\bm{k}_2)[1-f^0(\bm{k}_3)][1-f^0(\bm{k}_4)]\over [1-f^0(\bm{k})]} \cr
&\times (\delta_{\bm{k},\bm{k^\prime}} + \delta_{\bm{k}_2,\bm{k^\prime}} - \delta_{\bm{k}_3,\bm{k^\prime}} - \delta_{\bm{k}_4,\bm{k^\prime}}) \cr
&=  \sum_{i=1}^4 \mathcal{L}^{[i]}_{\bm{k},\bm{k^\prime}},    
\end{align}
and which can be written as the sum of four contributions. The explicit expressions for the diagonal ($\mathcal{L}^{[1]}$) and off-diagonal $(\mathcal{L}^{[2-4]})$ terms are calculated  according to Ref.\cite{Shahbazi15}, in the limit of the quasi-1D electron gas  model. Their expressions  given   in Appendix A are generalizations  at arbitrary energy distance  from  the Fermi level.

The electric current density along the chains resulting from a longitudinal thermal gradient $\nabla_aT$ and the induced electric field ${\cal E}_a $   in leading order is given by 
\begin{align}
\label{ }
j_a = &\  {2e\over L N_\perp d_\perp} \sum_{\bm{k}} v^a_{\bm{k}} f({\bm k}) \cr
\simeq &  {2e\over L N_\perp d_\perp} \sum_{\bm{k}} v^a_{\bm{k}} f^0({\bm k})[1-f^0({\bm k})] (\phi^{\cal E}_{\bm{k}} - \phi^T_{\bm{k}}),
\end{align}
where $v^a_{\bm{k}}$ is the carrier velocity along the $a$ direction.  Introducing the normalized deviations $\bar{\phi}^{\cal E}_{\bm{k}} = \phi^{\cal E}_{\bm{k}}/( e \beta v_{\bm{k}_F}^a   {\cal E}_a)$  and $\bar{\phi}^T_{\bm{k}}= \phi^T_{\bm{k}}/ [ {\beta}^2  k_B  v_{\bm{k}_F}^a (E_{\bm{k}} - \mu)\nabla_aT]$, which have the units of time, this expression can be recast in the  form 
\begin{equation}
\label{ }
j_a = K_{11} {\cal E}_a - K_{12}\nabla_aT,
\end{equation}
which leads in the absence of charge current ($j_a=0$) to the expression of longitudinal Seebeck coefficient $Q_a$, as the ratio
\begin{equation}
\label{ }
Q_a= {{\cal E}_a\over \nabla_aT} =  { K_{12}\over K_{11}}.
\end{equation}
Since the product $f^0[1-f^0]$ is strongly peaked at the Fermi level $E_{\bm{k}}-\mu \equiv E=0$, a Sommerfeld expansion of the matrix elements $K_{11}$ and $K_{12}$ yields the following expression  for the Seebeck coefficient,
\begin{align}
\label{thermopower}
Q_a =  {\pi^3\over 3} {k_B^2 T\over |e |} \Bigg\{ & \left[ -{ d\ln  \langle N(E,k_\perp)\rangle_{k_\perp}\over dE} - 2 {d\ln \langle v_{E,k_\perp}^a\rangle_{k_\perp} \over dE}\right]\cr
 &-   {\partial  \ln \langle\bar{\phi}_{E,k_\perp}\rangle_{k_\perp}\over \partial E}  \Bigg\}_{E=0}\cr
=  Q_a^0 + Q_a^c,
\end{align}
which can be separated into two contributions.   The first, noted $Q_a^0$,  is the sum of the two terms in brackets, which  corresponds to  the band  contribution. It is linked to the  energy derivatives of the density of states per spin, ${\langle N(E,k_\perp)\rangle_{k_\perp} (= \pi^{-1}\langle |\partial k/\partial E_{\bm{k}}| \rangle_{k_\perp}  }$),  and of the longitudinal velocity ${ \langle v_{E,k_\perp}^a\rangle_{k_\perp} ( = \hbar^{-1} \langle\partial  E_{\bm{k}} /\partial k \rangle_{k_\perp}  }$). Both quantities are averaged over the Fermi surface for a filling of one hole per dimer ($\langle \ldots \rangle_{k_\perp}= N_P^{-1}\sum_{k_\perp} \ldots$). The second contribution, $Q_a^c$, is associated to collisions; it is  proportional to the energy derivative $\langle\bar{\phi}'_{E=0,k_\perp}\rangle_{k_\perp}$ averaged over the Fermi surface for the normalized deviations $\bar{\phi}^{\cal E}=\bar{\phi}^T\equiv \bar{\phi}$, namely the   scattering time. The latter obeys   the single integral  equation
\begin{equation}
\label{BEQ}
{{\cal L}}\bar{\phi}_{\bm{k}} = \sum_{i,\bm{k'}}{\cal L}^{[i]}_{\bm{k},\bm{k'}} \bar{\phi}_{\bm{k'}}=1, 
\end{equation}
whose explicit expression is given in (\ref{LBEb}). Here $\bar{\phi}_{\bm{k}}\to \bar{\phi}_{E,k_\perp}$ can be expressed as a function of the energy distance from the Fermi surface and the angle parametrized by $k_\perp$. The expression (\ref{thermopower})  is reminiscent of the Mott formula for the Seebeck coefficient\cite{Mott69,Behnia15}. It should be stressed, however, that  the scattering  term results from the solution of the $\bm{k}$ dependent integral equation (\ref{BEQ}), which goes beyond the relaxation time approximation used for the Mott result\cite{Ziman72}. 
 \section{Renormalized umklapp vertex}
 \label{Model}
\subsection{The   quasi-one-dimensional electron gas model}
\label{RG}
The temperature variation of the momentum dependent umklapp vertex part entering the collision operator of the Boltzmann equation (\ref{CollOp}) is calculated using the renormalization group technique in the framework of the quasi-one-dimensional electron gas model. In the model  the longitudinal part of  lattice model for the hole spectrum  $E_{\bm{k}}$ in (\ref{Ek}) is linearized with respect to the 1D Fermi points $\pm k_F$. This gives
\begin{equation}
\label{linE}
E_{\bm{k}}-\mu \approx \epsilon_{\bm{k}}^p=-\hbar v_F (pk-k_F) + \epsilon(k_\perp),
\end{equation} 
where $p=\pm$ refers to right and left moving carriers along the chains and $v_F = (t^2-\delta t^2)a/(\hbar\sqrt{2t^2+2\delta t^2})$ is the longitudinal Fermi velocity. According to band calculations,  the hopping integrals will be fixed at  $t/k_B= 2700$K and  $t_\perp/k_B= 200$K as typical figures for hopping integrals in compounds like the Bechgaard salts. A second harmonic is added to the transverse tight-binding spectrum which acts  as an anti-nesting tuning parameter $t_\perp'\ll t_\perp$.  Anti-nesting is considered as the main parameter simulating the pressure in  the model.

Particles interact through three coupling constants defined on the warped Fermi surface sheets $\bm{k}_F^p(k_\perp) =(k_F^p(k_\perp),k_\perp)$, as parametrized by $k_\perp$ from the condition  ${\epsilon^p(\bm{k}_F^p)=0}$ (see the top panel  of the Fig.~\ref{dPhidEht}). These are  the backward and forward scattering amplitudes $g_1(\boldsymbol{k}_{F,1}^-,\boldsymbol{k}_{F,2}^+;\boldsymbol{k}_{F,3}^-,\boldsymbol{k}_{F,4}^+)$ and $g_2(\boldsymbol{k}_{F,1}^+,\boldsymbol{k}_{F,2}^-;\boldsymbol{k}_{F,3}^-,\boldsymbol{k}_{F,4}^+)$,   and the longitudinal umklapp scattering $g_3(\boldsymbol{k}_{F,1}^p,\boldsymbol{k}_{F,2}^p;\boldsymbol{k}_{F,3}^{-p},\boldsymbol{k}_{F,4}^{-p})$.  All couplings are normalized by   $\hbar \pi v_F$ and develop from renormalization a momentum dependence on  three independent transverse momentum variables. 

We will follow previous works\cite{Bourbon09,Sedeki12,Shahbazi15} and fix the bare initial repulsive values of the couplings  consistently with different experiments and band calculations. Thus for the bare backward scattering, by taking $g_1\approx 0.32$, one can reasonably account for the observed temperature dependent enhancement of uniform susceptibility \cite{Wzietek93}. For the bare longitudinal umklapp term $g_3$,  its bare amplitude is non-zero but very weak, owing to the small dimerization of the organic stacks that introduces some  half-filled character to the band. This yields to  $g_3 \approx g_1 (2\delta t/t)$,  as a result of  the modulation $\delta t$ of longitudinal hopping integrals responsible for the dimerization gap\cite{Barisic81,Penc94}.  According to band calculations at low pressure\cite{Ducasse86},   ${\delta t/t \approx 0.05 ... 0.1}$, suggesting to take $g_3\approx 0.025 $ in the following calculations.  From these figures,  the amplitude of the bare forward scattering can be finally  adjusted   to the value $g_2 \approx 0.64$, in order to  get from the   low $t_\perp'$ RG calculations (see Fig. \ref{Phases})     the right order of magnitude  for the observed SDW scale, namely $T_{\rm SDW}\sim 10$K   for (TMTSF)$_2X$ at ambient pressure\cite{Jerome82}.  As a function of $t_\perp'$, the application of the RG generates a phase diagram compatible with the experimental situation\cite{Jerome82,DoironLeyraud09}. There is nothing special in the above choice. Actually at small umklapp, it exists a whole range of reasonable coupling parameters  that would  yield a phase diagram comparable to Fig.~\ref{Phases} and then to similar results for the Seebeck coefficient.

 As it will discussed in more details in Sec.\ref{Fabre}, one can extend the analysis to the more correlated sulfur based Fabre salts series (TMTTF)$_2X$  characterized by smaller band  parameters and stronger umklapp scattering owing  to a larger dimerization of the organic stacks.\subsection{Renormalization group results}
\label{RGB}
The RG approach to the above quasi-1D electron gas model has been described in detail in previous works\cite{Nickel06,Bourbon09,Sedeki12,Fuseya12,Shahbazi15}. In essence, it consists in the segmentation of infinitesimal energy shells on either side of the Fermi sheets into $N_P$ patches, whose internal transverse momentum integration in the loop calculations, leads to as many $k_\perp$ values. Successive integrations of electronic degrees of freedom on these shells from the  (Fermi)  energy cutoff $E_F/k_B[= \pi t /(2k_B\sqrt{2})]\equiv 3000$K down to zero at the Fermi surface result in the flow of the coupling constants toward their momentum dependent values as a function of temperature. This is carried out until a singularity is reached in the coupling constants which signals  an instability of the electron gas against the formation of a broken symmetry state at a given temperature. 

\begin{figure}
\includegraphics[width=8cm]{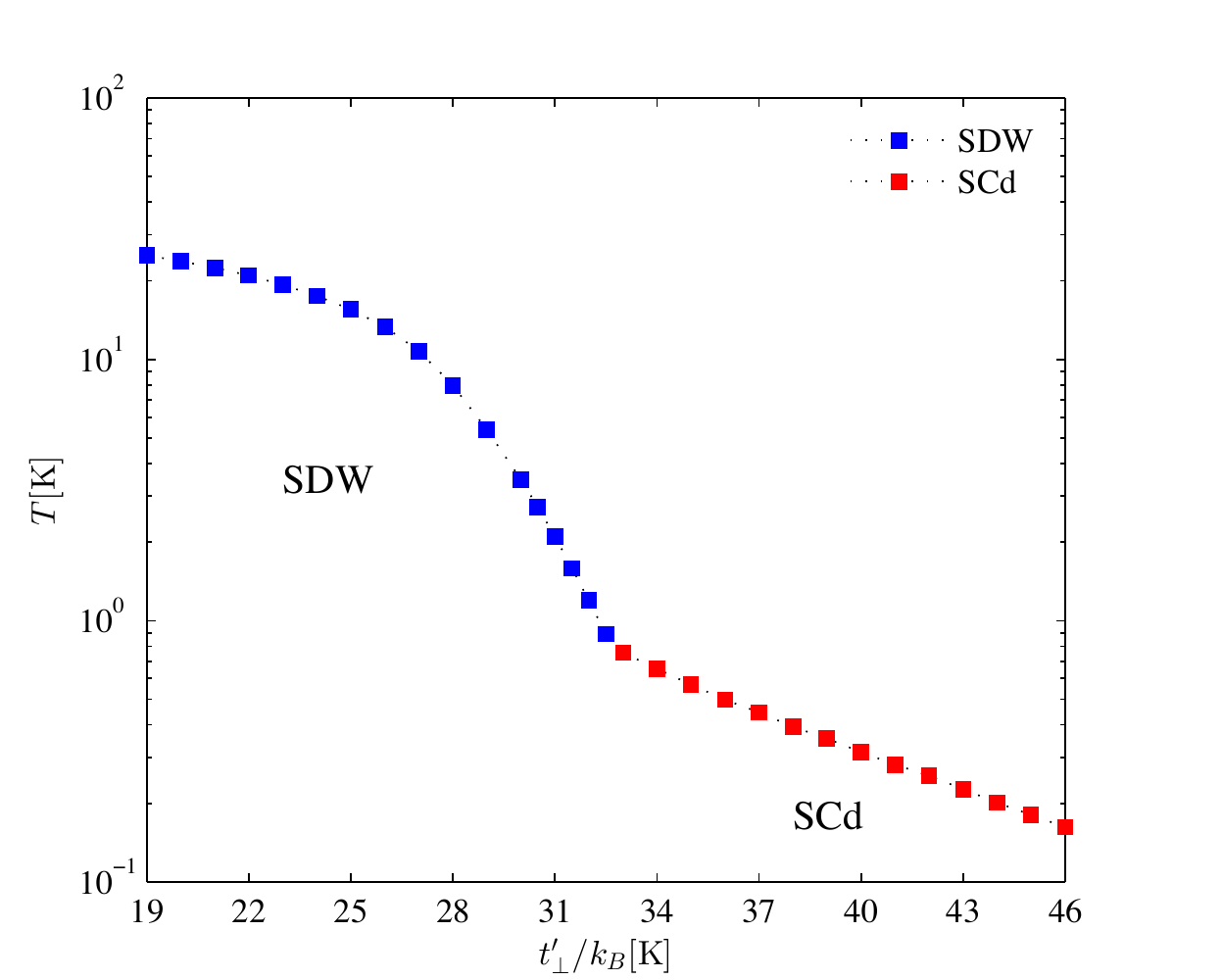}
\caption{Renormalization group results for  the phase diagram of the quasi-1D electron gas model  as a function of   the antinesting tuning parameter $t_\perp'$ and for the model parameters specified in Sec. \ref{RG}.}
\label{Phases}
\end{figure}

For the  repulsive sector with this (TMTSF)$_2X$ model parameters, this can occur  in either   SDW or d-wave SC (SCd) channel depending on the amplitude of antinesting $t_\perp'$. The characteristic sequence of instabilities  obtained for $N_P=60$~patches is reviewed  in  Figure~\ref{Phases}\cite{Shahbazi15}. At  relatively low nesting deviations the magnetic  scale $T_{\rm SDW}$ dominates; it drops with $t_\perp'$  down to  the critical value $t_\perp'^*$ where instead of a plain quantum critical behaviour for antiferromagnetism  for which  $T_{\rm SDW}$   would reach zero, the  ending of $T_{\rm SDW}$ gives rise to an SCd instability  at its maximum $T_c$. The latter then  steadily falls off  with further increasing $t_\perp'$.

The normal phase we are interested in for the Seebeck coefficient is characterized by spin fluctuations. This is of course found in the  SDW sector of the phase diagram where the SDW  susceptibility  $\chi_{\rm SDW}(\bm{q}_0)$  at   the best nesting wave-vector of $\epsilon_{\bm{k}}^p$ at $\bm{q}_0=(2k_F,\pi/d_\perp)$, develops  a singularity  at  $T_{\rm SDW}$. In the   SCd sector, an enhancement, though non singular, is still present. It takes    
  the form of a Curie-Weiss temperature profile $\chi_{\rm SDW} \sim (T+\Theta)^{-1} $ over a large temperature domain above $T_c$ ($\Theta \ge 0$). The enhancement is quantum critical at $t_\perp'^*$ ($\Theta =0$) and  then decays with the decrease of $T_c$ and the rise of the Curie-Weiss scale $\Theta$ along the antinesting axis \cite{Bourbon09}.

\begin{figure}
\includegraphics[width=4.2cm]{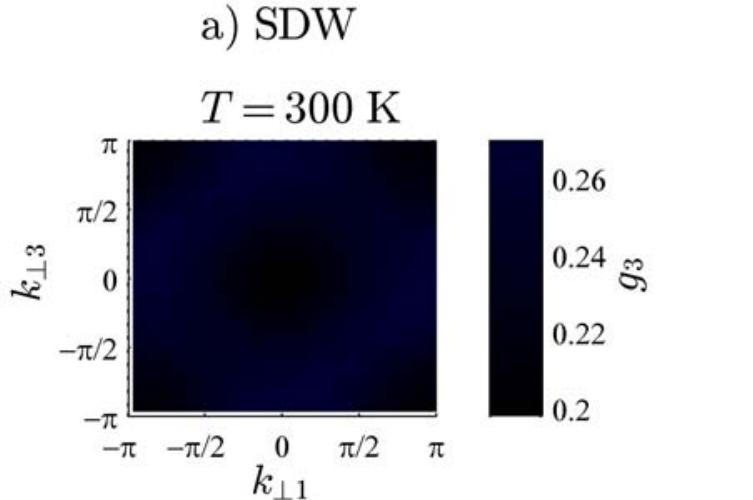}
 \includegraphics[width=4.2cm]{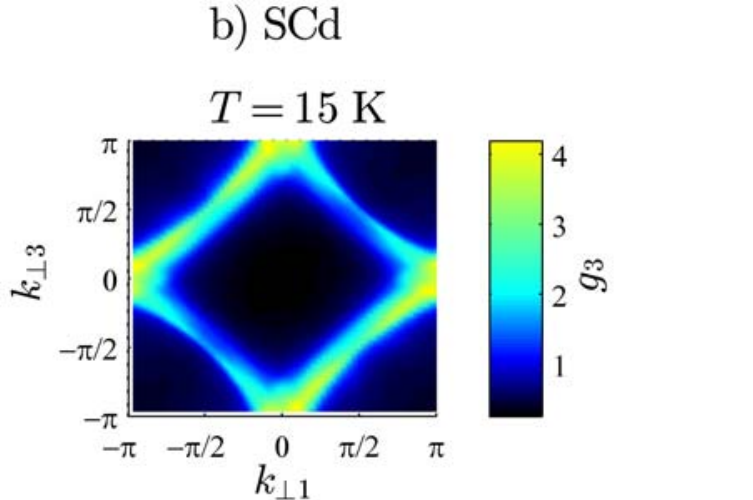}  \\ 
    \includegraphics[width=4.2cm]{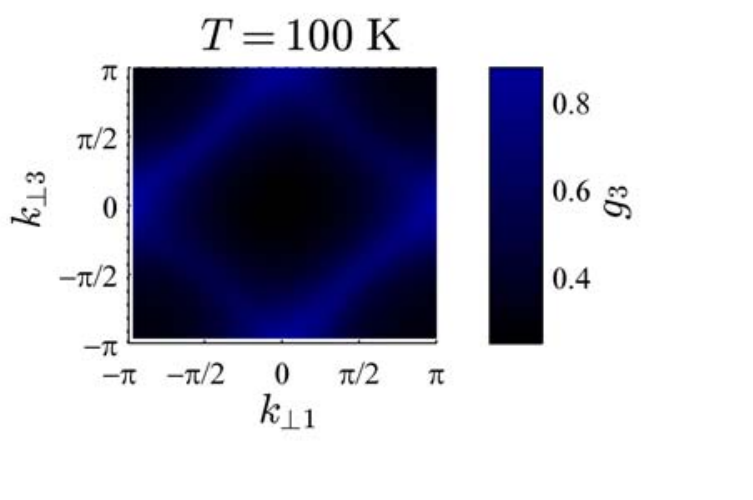}
     \includegraphics[width=4.2cm]{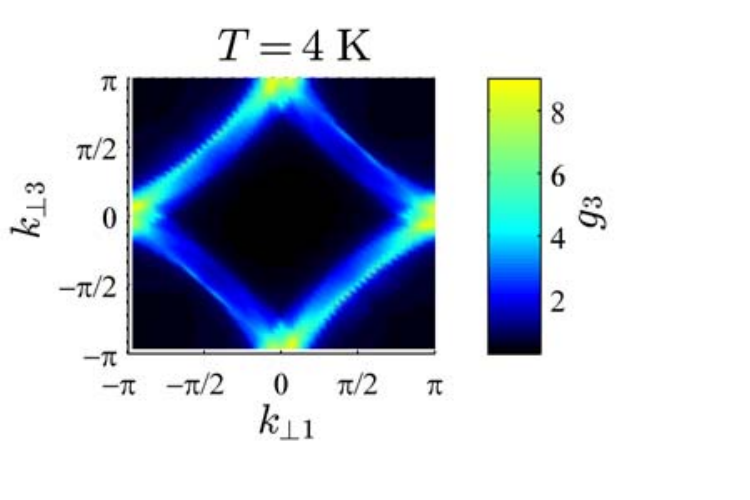} \\
      \includegraphics[width=4.2cm]{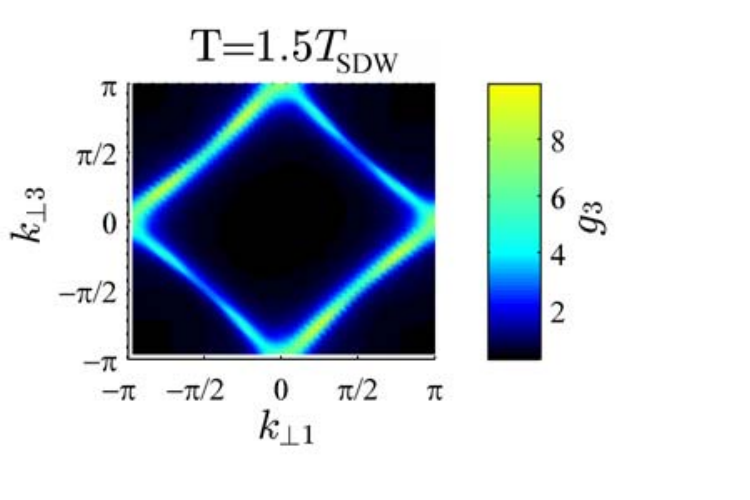}
     \includegraphics[width=4.2cm]{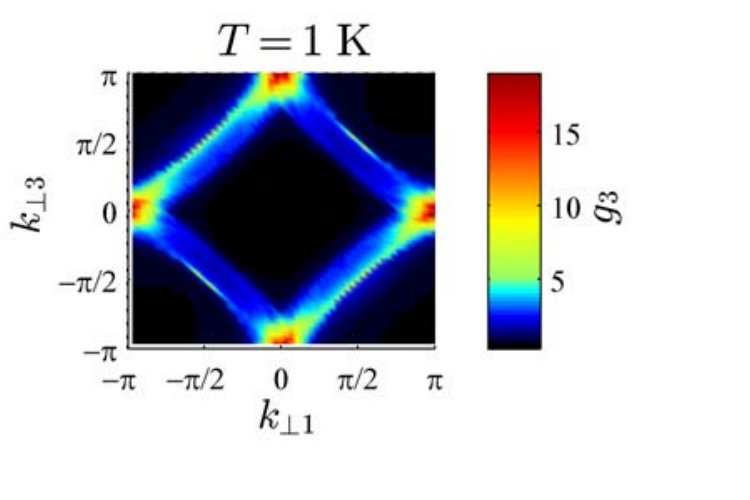} 
  \caption{Renormalized umklapp scattering amplitude $g_3(k_{\perp1},-k_{\perp1}; k_{\perp3},-k_{\perp3})$, projected in the $(k_{\perp1},k_{\perp3})$ plane at  different temperatures for the metallic phase model parameters specified in Sec.~\ref{RG}. (a): SDW, $t_\perp'/k_B= 25$K $(<t_\perp'^*/k_B)$; (b) : SCd, $t_\perp'/k_B= 35$K $(>t_\perp'^*/k_B)$. \label{g3}}
\end{figure}

These short-range SDW correlations of  the metallic phase are directly related to the enhancement of umklapp scattering entering   the collision operator of the Boltzmann equation. In Figure~\ref{g3}, we show the temperature and momentum dependence of $g_3$ on the Fermi surface, as projected in  the$(k_{\perp1},k_{\perp3})$ plane when $k_{\perp1}=-k_{\perp2}$ and $ k_{\perp3}=-k_{\perp4}$. On the SDW side, the top panel of the Figure~\ref{g3}-a refers to the high  temperature range ($T>t_\perp$) which shows no structure  
in the transverse momentum plane for $g_3$, indicating that SDW correlations are essentially 1D in character  and confined along  chains. As temperature is lowered below $t_\perp$, transverse short-range order starts to develop, as shown by more intense scattering along the lines $k_{\perp1} = k_{\perp3} \pm \pi $ {($d_\perp=1$)}. This is in accordance with the transverse momentum transfer associated with  the best nesting wave-vector $\bm{q}_0$ of   the spectrum (\ref{linE}). When   the lowest temperature is reached, peaks of stronger intensity appear on the corners at $k_{\perp1,3} =0,\pm \pi $ and at the best nesting points $\pm \pi/4, \pm 3\pi/4$ of the spectrum (\ref{linE}). These refer to warmer regions of scattering on the Fermi surface at the approach of the critical domain of the SDW instability.

On the SCd side of the phase diagram, in  Fig.~\ref{g3}~(b), we see a   pronounced   but non singular   anisotropic increase of umklapp scattering; peaks are    confined around  $k_{\perp1,3}=0,\pm \pi$  on  the Fermi surface, where  enhanced scattering is found as the temperature is lowered. This enhancement occurs  despite  the flow of coupling constants    towards a SCd fixed point indicative of a positive interference between both  instabilities.   This increase goes hand in hand with the one of SDW correlations in this temperature region, which are directly involved in the mechanism of d-wave Cooper pairing\cite{BealMonod86,Caron86,Emery86}.  By increasing $t_\perp'$ further, one can show that, although the same anisotropy of umklapp  enhancement persists,  its amplitude scales down with the reduction of $T_c$.  

The consequence of this anisotropic growth of umklapp scattering  on the temperature dependence of the Seebeck coefficient will be analyzed next.

\section{Numerical results for the Seebeck coefficient} 
\subsection{High temperature domain}
\begin{figure}
\includegraphics[width=8cm]{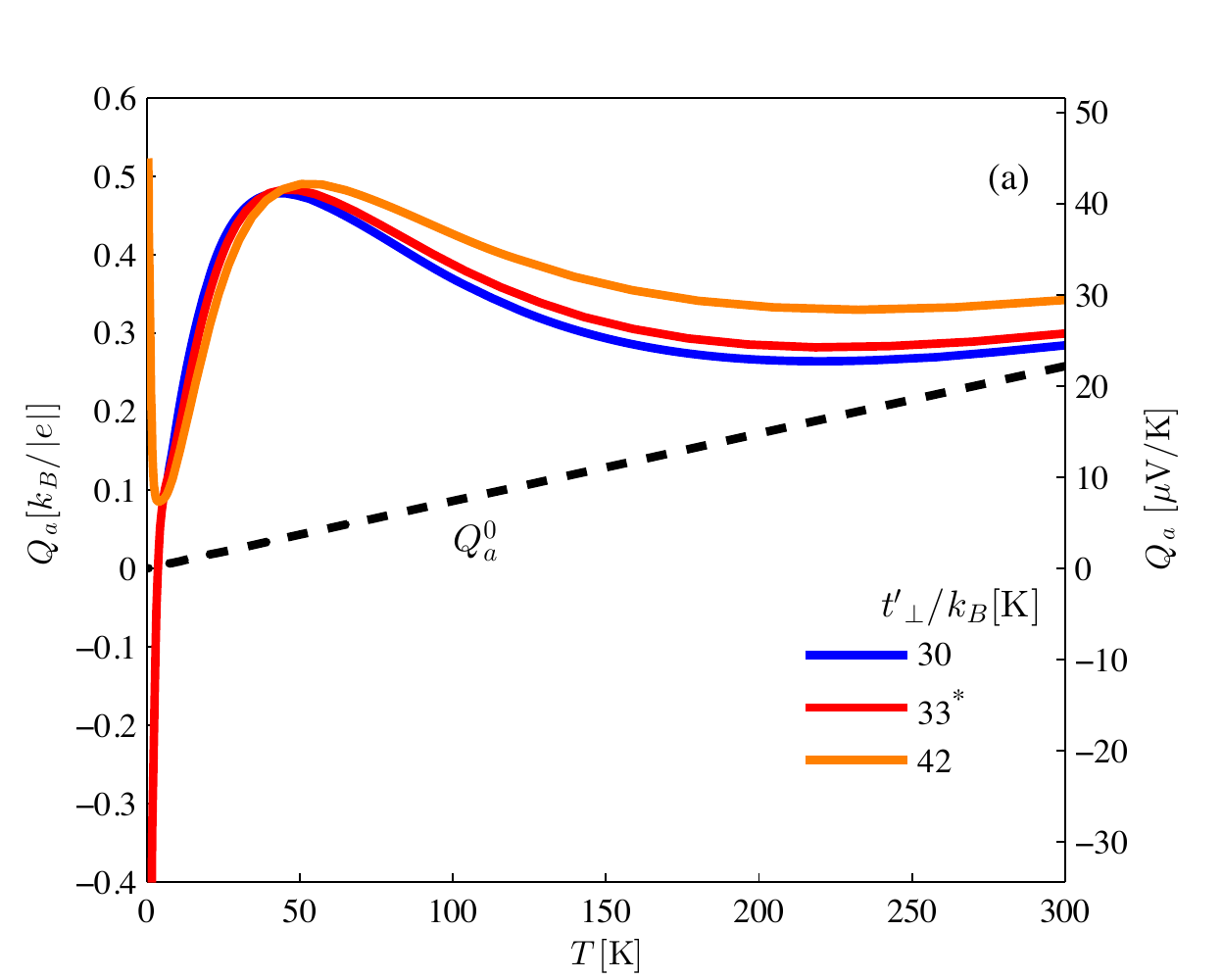} \\ 
\includegraphics[width=8cm]{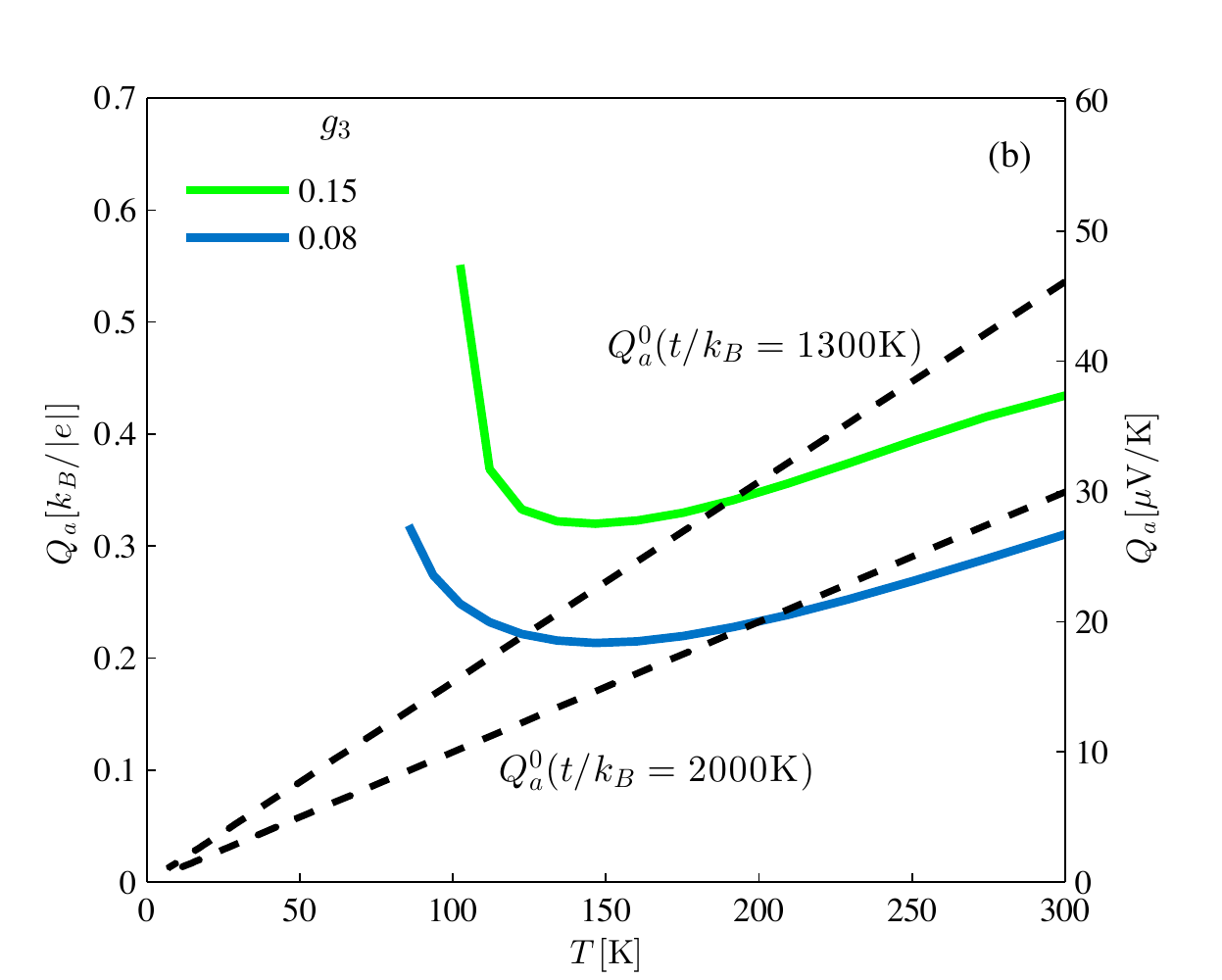}
\caption{ The longitudinal Seebeck coefficient  as a function of temperature for (a) different values of antinesting $t_\perp'$ in the metallic phase and (b) for model parameters in the more correlated case with stronger umklapp scattering and lower amplitudes of  hopping integrals ($t_\perp'/k_B= 15$K, $t_\perp/k_B=100$K, see Sec.\ref{Fabre} on (TMTTF)$_2X$ salts). The dashed lines gives the band contribution $Q_a^0$  of Eq.~\ref{thermopower}  for constant relaxation time in energy. }
\label{Qht}
\end{figure}

\begin{figure}
\includegraphics[width=8cm]{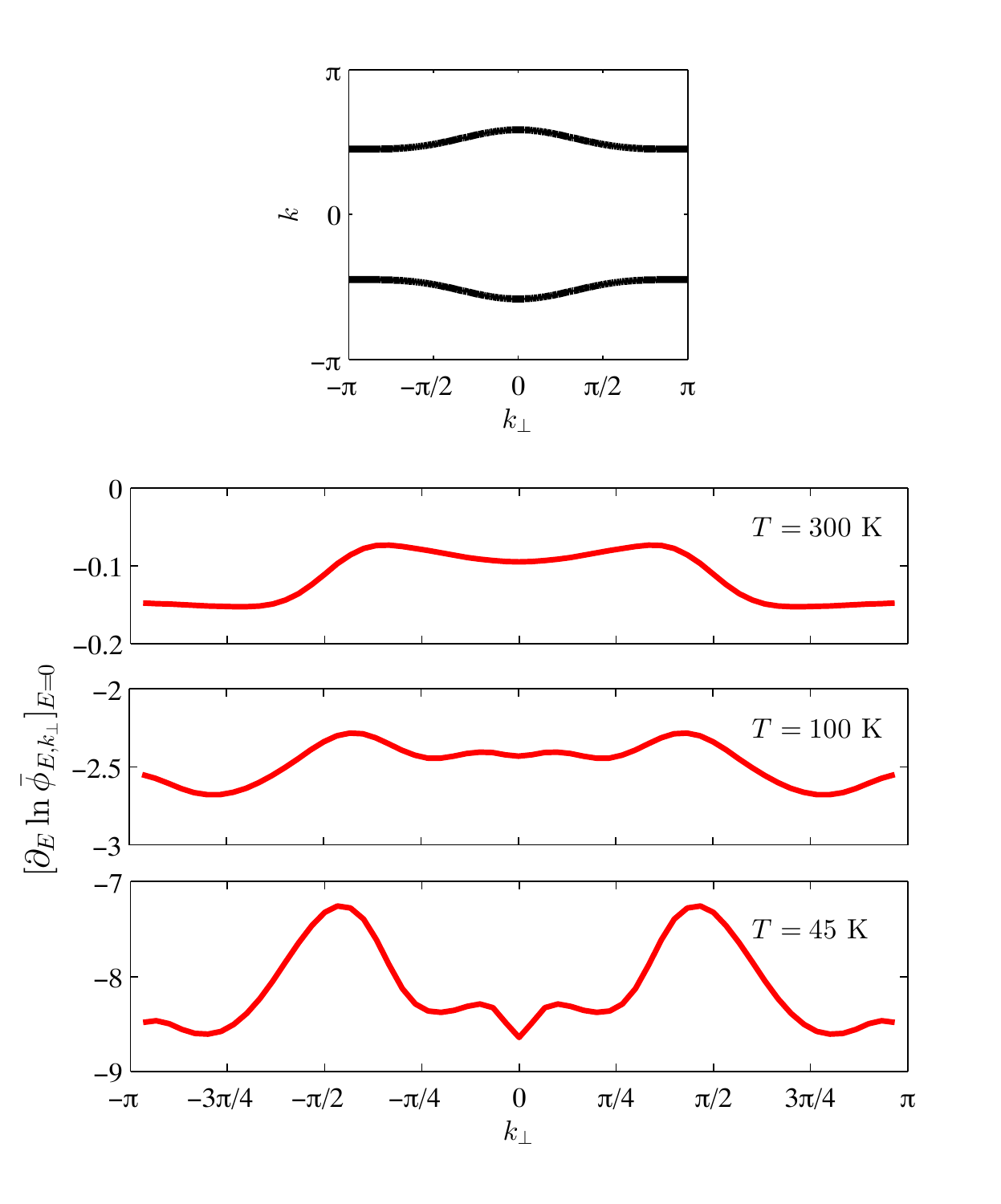}
\caption{ Open Fermi surface of the quasi-1D electron gas model (top panel) and typical variations of the scattering time along the Fermi surface for different high temperatures ($t_\perp'=t_\perp'^*)$. }
\label{dPhidEht}
\end{figure}

The   temperature dependence  of the Seebeck coefficient (\ref{thermopower}), as  obtained from the numerical solution of  (\ref{LBEb}) for the (TMTSF)$_2X$ model parameters, is shown  in Fig.~\ref{Qht} (a)  in the whole temperature  interval of interest. By comparing the  amplitude of the two contributions to the Seebeck coefficient in (\ref{thermopower}), we observe that apart from the high 1D temperature region the amplitude of the last term related to scattering  dominates the band  contribution $Q_a^0$  (dashed line of Fig.~\ref{Qht} (a)) over most of  the temperature interval.
 This gives rise to a shallow minimum  for the Seebeck coefficient below which the  normalized energy derivative $\bar{\phi}'_{E=0,k_\perp}/\bar{\phi}_{E=0,k_\perp}$ of the  normalized scattering time on the Fermi surface grows in importance, as meant  in the lower panels of Fig.~\ref{dPhidEht}. The  derivative is negative and according to (\ref{thermopower}), it gives a positive   $Q^c_a$, as normally expected for holes carriers whose  velocity and scattering time    decrease  with increasing energy.
 
  By lowering temperature the scattering time derivative  gains in amplitude and develops,  like $g_3$ of Fig.~\ref{g3},   anisotropy over the Fermi surface with maximums at $k_\perp =0, \pm \pi$ and $\pm \pi/4, \pm 3\pi/4$.
This leads  to a smooth increase of the Seebeck coefficient that levels off at a maximum value around  the antinesting $t_\perp'$ scale. This is followed in Fig.~\ref{Qht}~(a)  by a rapid drop at lower temperature which is nearly linear; it evolves toward  anomalous features  in the amplitude or the sign  of the Seebeck coefficient  depending on the distance to the  critical value $t_\perp'^*$ in the phase diagram. This will be discussed in more details below. 

\subsection{Low temperature domain and quantum criticality}
\begin{figure}
\includegraphics[width=8cm]{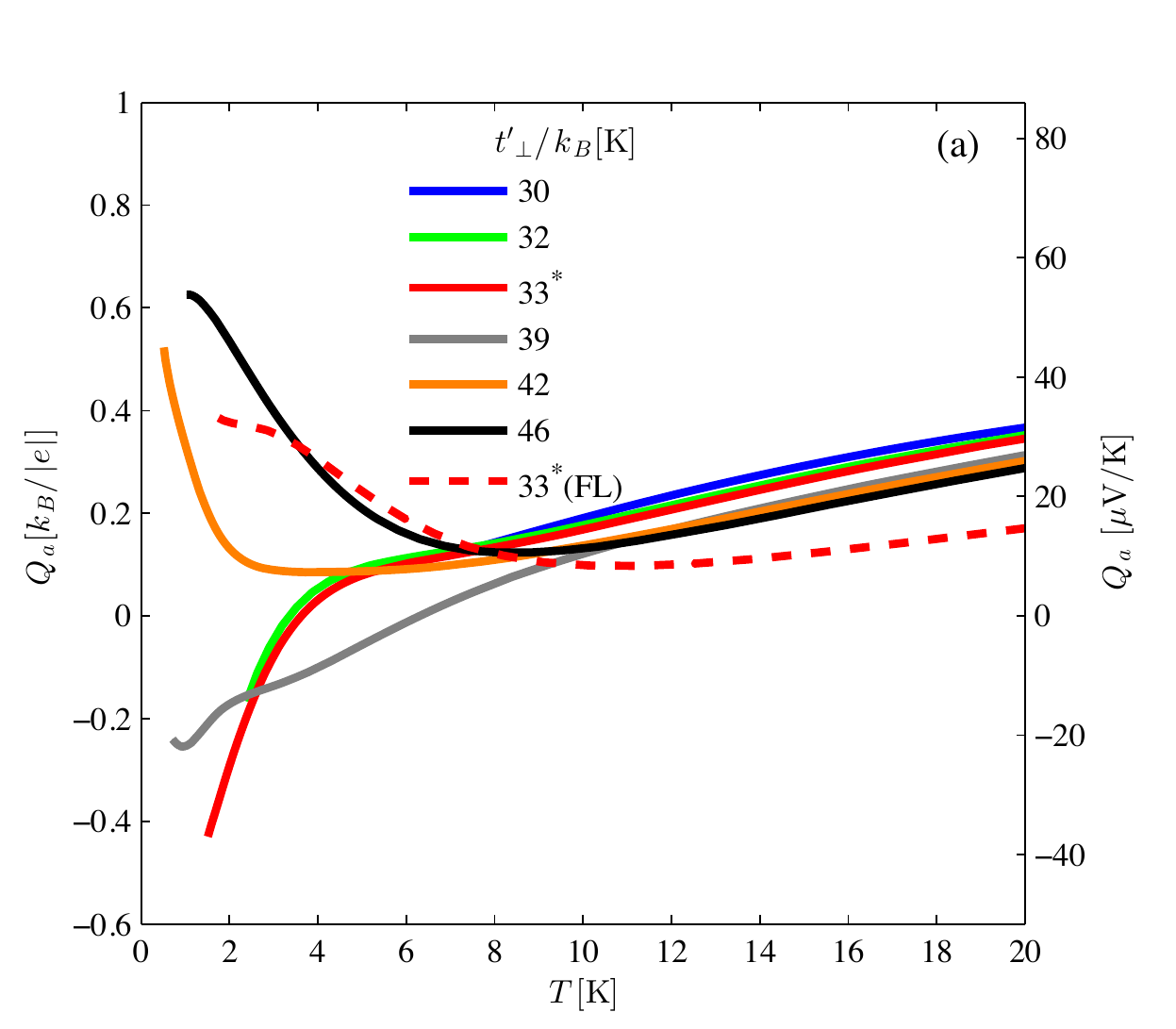}\\
 \includegraphics[width=8cm]{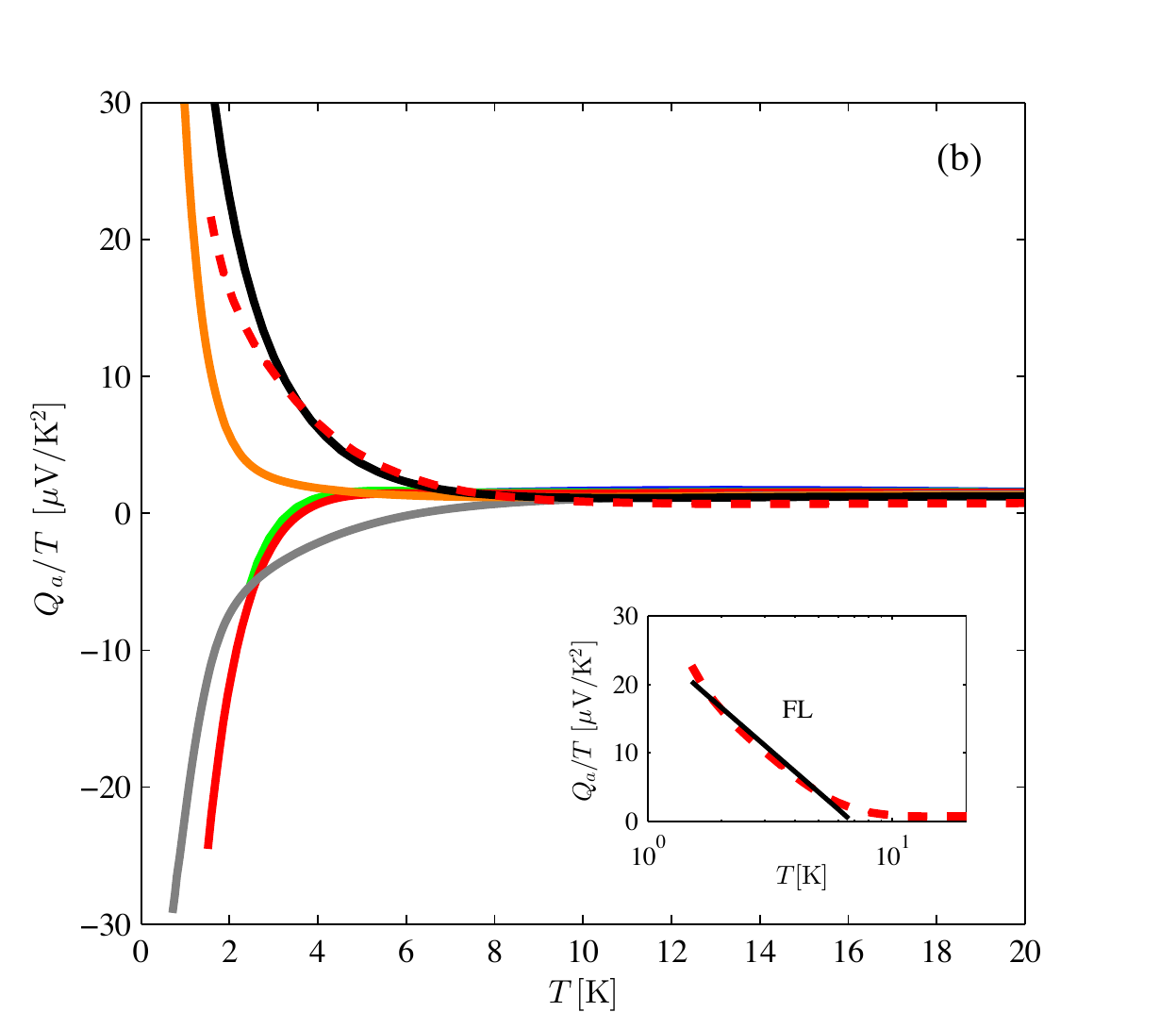}  
  \caption{ (a) The longitudinal Seebeck coefficient $Q_a$ and (b) the ratio $Q_a/T$, as a function of $T$ at low temperature  and  different values of antinesting $t_\perp'$. The value with  asterisk stands for the critical $t_\perp'^*$. The dashed line corresponds to the Fermi liquid limit using a momentum and temperature independent $g_3$ ($=0.025$) at $t_\perp'^*$. The inset of part (b) displays the enhancement on a logarithimic temperature  scale. The continuous line refers to $Q_a/T \sim \ln T.$   \label{Q}}
\end{figure}

\begin{figure}
\includegraphics[width=8cm]{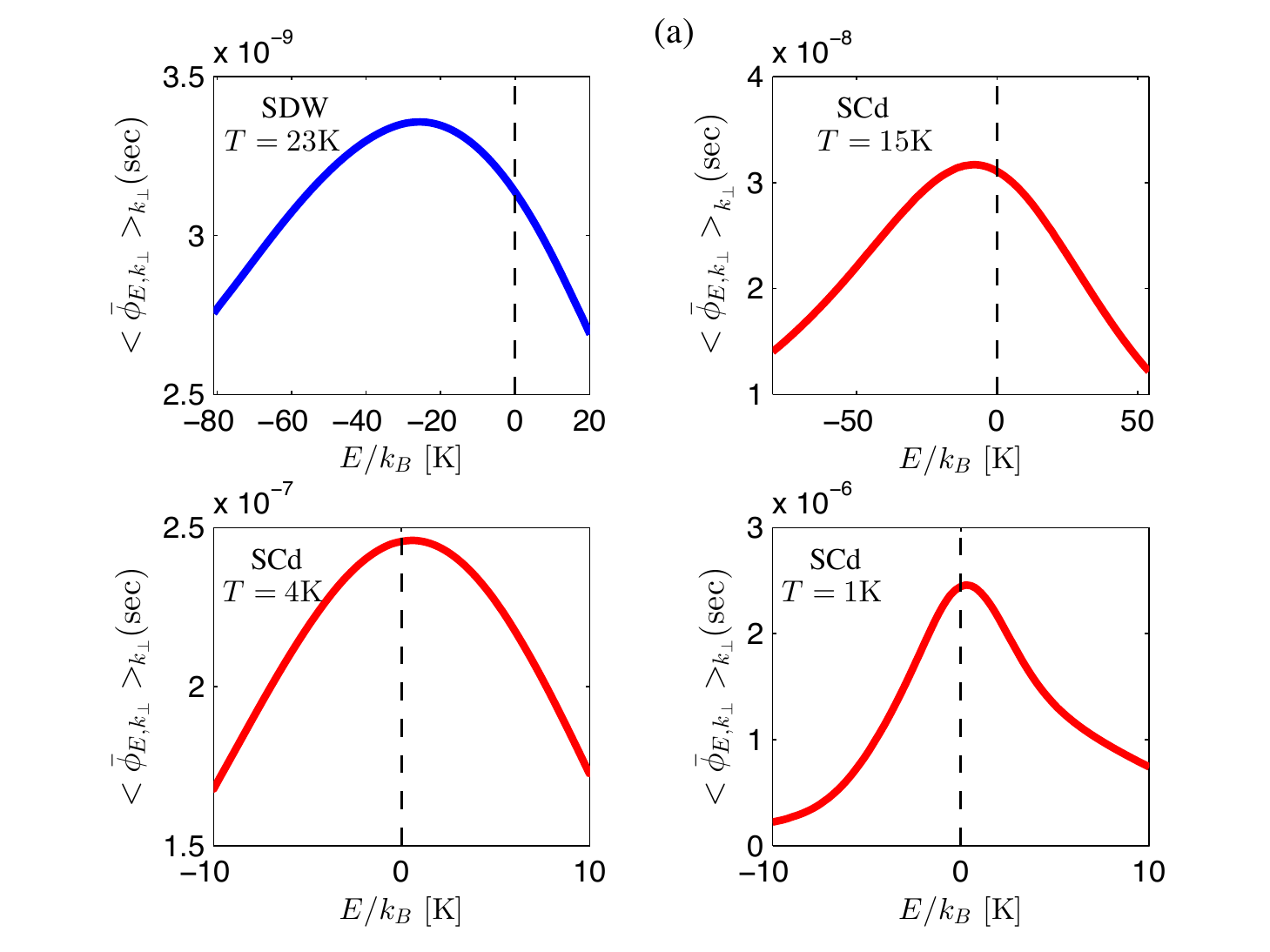}\\
\includegraphics[width=8cm]{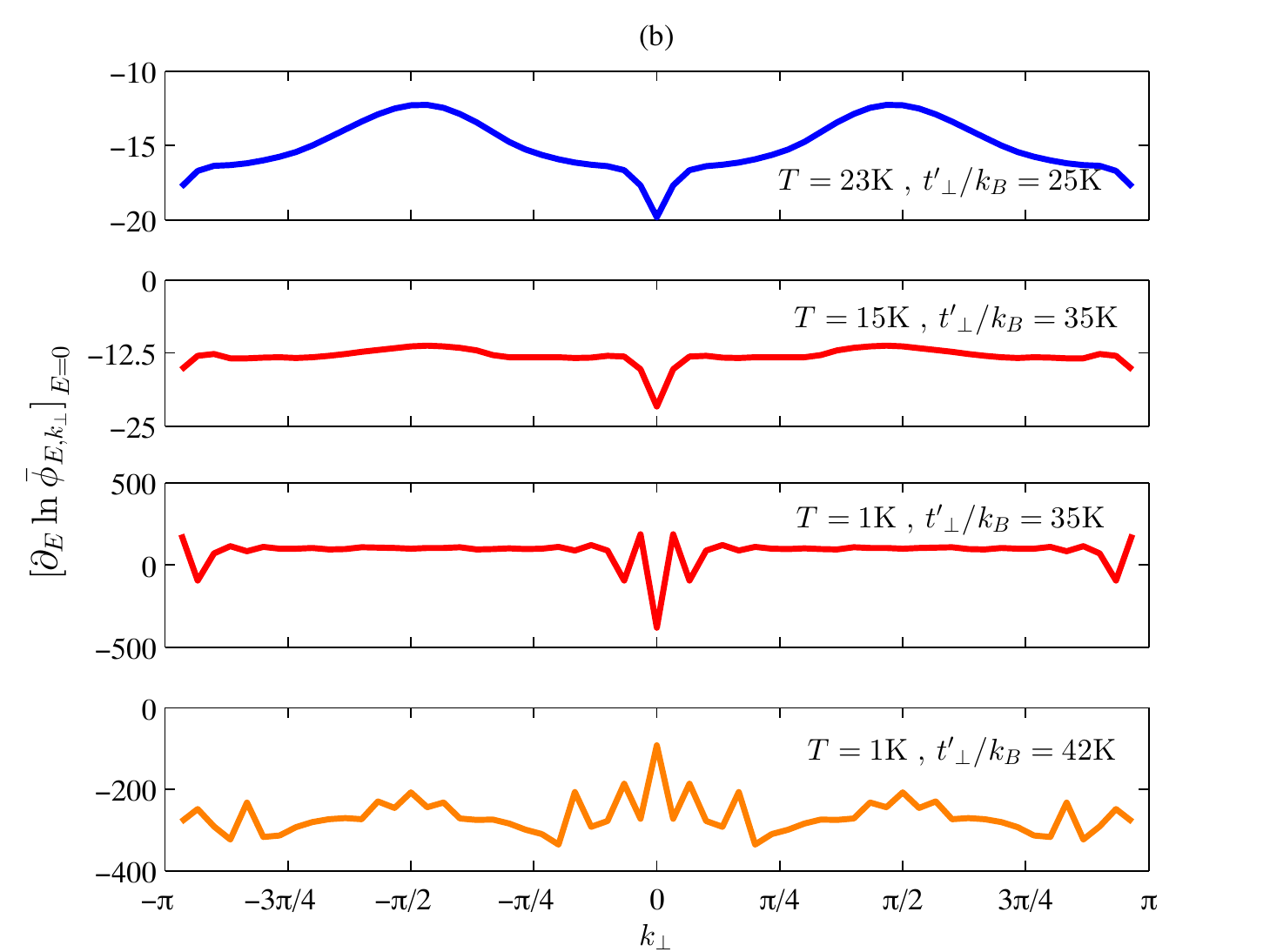}
  \caption{ (a) Variation of the normalized scattering time as a function  of energy near the Fermi level at low temperature. Here $t_\perp'/k_B= 25$K and $35$K for the blue and red curves, respectively. (b) The anisotropy of the  Seebeck coefficient along the Fermi surface  at low temperature for different $t_\perp'$. \label{dphidE}}
\end{figure}

The results for the Seebeck coefficient in the  metallic low temperature  part of the phase diagram are presented in Fig.~\ref{Q}~(a) for different values of the antinesting parameter $t_\perp'$. On the SDW side, for  $t_\perp'$ relatively well below the critical $ t_\perp'^*/k_B$ (= 33K), the  decrease of the Seebeck coefficient with lowering temperature  is nearly linear as indicated  by the constant ratio $Q_a/T$ in Fig.~\ref{Q} (b) when the temperature is lowered. Here the slope for $Q_a$ is steeper than for the band contribution  $Q_a^0$  (dashed line Fig.~\ref{Qht} (a). The dominant  contribution to the Seebeck coefficient is coming from $Q_a^c$ which is positive,   resulting from a peak in the  energy dependent quasi-particle scattering time located on the occupied side of the Fermi level at $E<0$, as shown in the first top left panel of Figure~\ref{dphidE}~(a) above the SDW instability. It is worth noting that in these metallic conditions of the SDW state,  the calculated scattering time at the Fermi level ($\sim 10^{-9}$~sec) is significantly larger than the one found  from a Drude theory of the conductivity of the Bechgaard salts above  the SDW state\cite{Dressel96}(see the note in Ref.\cite{NoteSeebeck1}).

By raising $t_\perp'$, the temperature  scale   $T_{\rm SDW}$ in  Figure~\ref{Phases} decays and at the approach of  $t_\perp'^*$ from below, the Seebeck coefficient develops an anomalous   enhancement  that is  opposite in sign. This is depicted by the green lines 
in Fig~\ref{Q}.  The effect is reinforced when the  electron system ultimately enters the SCd domain at $t_\perp'^*$ where $T_c$ is maximum.  This indicates that the  collision   contribution $Q^c_a$  is still negative or electron like in character and that it exceeds  $Q_a^0$ in amplitude. The sign reversal of the Seebeck coefficient refers to an increase of the scattering time with energy  and then to a different asymmetry in the quasi-particle resonance peak of $\langle \bar{\phi}_{E,k_\perp}\rangle$. According to  Fig~\ref{dphidE}~(a), when the temperature is lowered, the latter  is   shifted from the occupied to  the unoccupied side just above the Fermi level  at $E>0$.  As for the anisotropy profile of $\bar{\phi}'/\bar{\phi}$ over the Fermi surface, the third panel of Fig.~\ref{dphidE}~(b) reveals that this  electron like component of the Seebeck coefficient comes in large part from the cold regions of scattering, namely, away from  the warmer spots  centered in  ${k_\perp=0}$ and $\pm \pi$ in the SCd sector [c.f. Fig.~\ref{g3}~(b)]. In  the latter  regions  large oscillations of $\bar{\phi}'/\bar{\phi}$ between   positive and negative values    tend to average out their contributions to a net positive contribution to the Seebeck coefficient.
It is worth noting that the  change of sign of the Seebeck coefficient, obtained by tuning $t_\perp'$ across $ t_\perp'^* $, occurs in the metallic state, that is in  the absence of any reconstruction of the Fermi surface. 

Further above  $ t_\perp'^* $, the negative enhancement of the Seebeck coefficient weakens  and finally transforms into a positive upturn, as shown in Fig.~\ref{Q}. The latter  is consistent with a quasi-particle resonance in the scattering time  whose peak shifts back below the Fermi level with a negative slope in $\langle \bar{\phi}'_{E=0,k_\perp}\rangle_{k_\perp} $, as shown in the lowest panel of Fig.~\ref{dphidE} (b).  This   contributes positively to $Q_a^c$. At sufficiently large $t_\perp'$, namely when nesting alterations become large,  the $g_3$ coupling renormalizes less and less with a concomitant weaker anisotropy. In these conditions  the Seebeck coefficient should tend  to that of a Fermi liquid. This is confirmed when one  imposes  a momentum and temperature independent $g_3$ in the calculations  of the scattering time in (\ref{LBEb}) which simulates the conditions of a Fermi liquid. This is shown by the dashed lines in   Figure~\ref{Q}. 
It is worth noting that the Fermi liquid result for the quasi-1D electron gas model differs from the  linear-$T$ band term $Q^0_a$.  The corresponding ratio $Q_a/T$ in Fig.~\ref{Q} (b) displays  a low temperature variation roughly congruent with a logarithmic enhancement.  This has to be related to the fact  with the fact that for a  quasi-1D Fermi liquid, the scattering time  is energy dependent and goes like $\sim E^2\ln E$ with logarithmic corrections; it is also  asymmetric with respect to the Fermi level due to  the presence of antinesting\cite{Gorkov98,Shahbazi15}. According to Figure~\ref{Q} (b), the effect of renormalized umklapp leads to enhancements slightly to logarithmic corrections expected by previous predictions near a quantum critical point\cite{Paul01}.

\begin{figure}
\includegraphics[width=10cm]{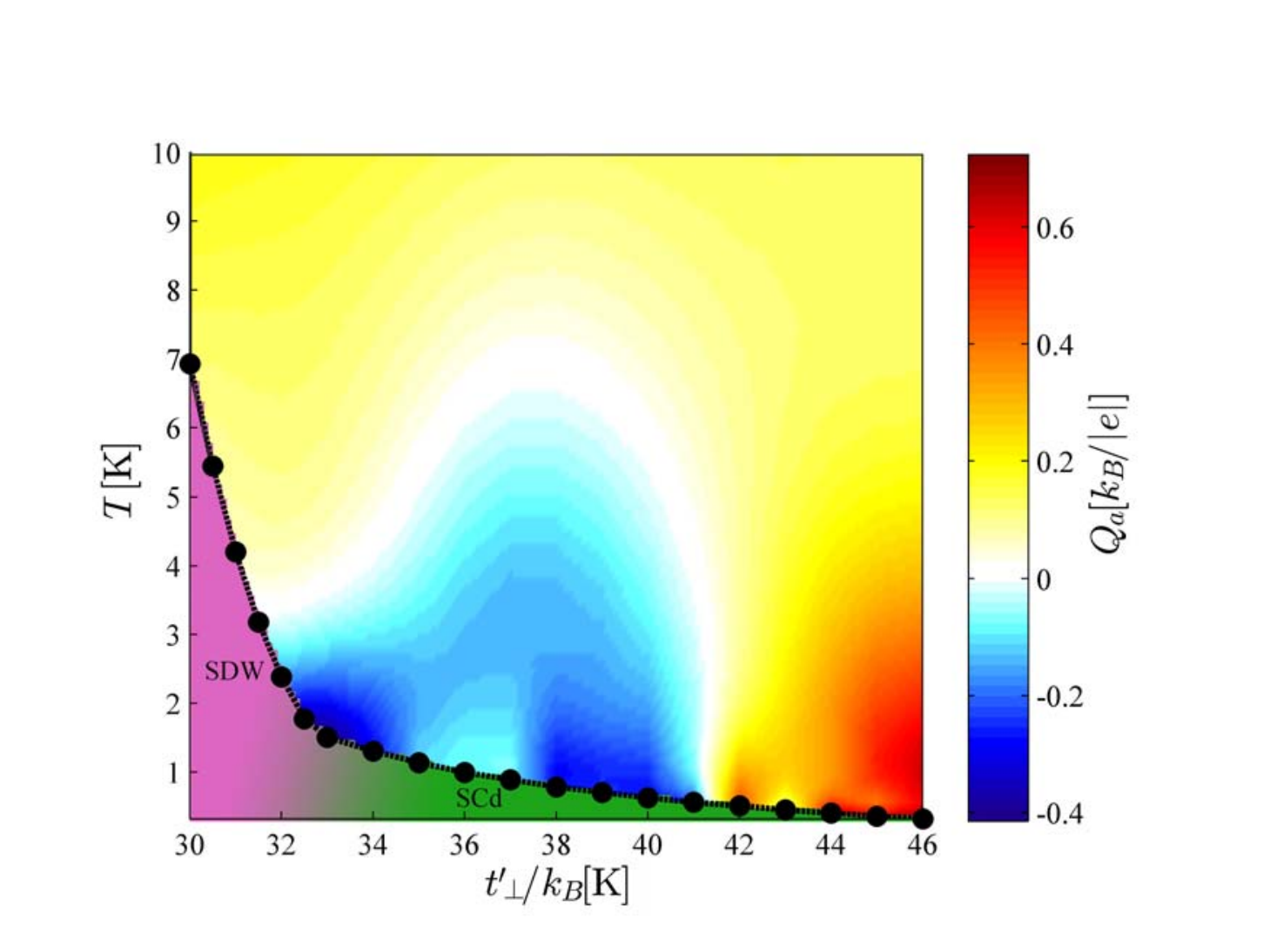} \caption{Amplitude of the Seebeck coefficient at low temperature as a function of antinesting. }
\label{Qint}
\end{figure}

Following the example of resistivity\cite{Shahbazi15}, one can define  from $t_\perp'^*$ a characteristic zone of influence of quantum criticality where an  anomalous sign of Seebeck coefficient is found.This is portrayed in Figure~\ref{Qint}. As pointed out previously\cite{Shahbazi15,Bourbon09,Sedeki12,Duprat01},  $t_\perp'^*$ defines a quantum critical point where the entanglement or mutual reinforcement between SDW and SCd instabilities is  the strongest, $T_c$  the highest, and where spin fluctuations are quantum critical down to $T_c$\cite{Bourbon09}. This is apparently responsible for the electron type  asymmetry in the energy dependence of electron-electron scattering time and therefore for the sign reversal of the  Seebeck coefficient. 

\section{Comparison with  experiments  in low dimensional organic conductors}
We now turn to the comparison of the above results with experiments. In this matter, it is  instructive to  first examine the Seebeck coefficient   for some members of the  (TMTTF)$_2X$ series, the so-called Fabre salts, which are known to exhibit a more correlated normal phase than the Bechgaard   salts  in normal pressure conditions as a result of stronger umklapp scattering\cite{Emery82,Bourbon08}. The weak coupling RG can be used to compute the flow of umklapp scattering down to the approach of the Mott strong coupling region. After this incursion in the physics of the Fabre salts, we  then proceed to the discussion of the Seebeck coefficient experiments in  the Bechgaard salts in the light of the results of the present calculations. 
\subsection{The  Fabre salts (TMTTF)$_2X$}
\label{Fabre}
 The Fabre salts with $X=$ PF$_6$, AsF$_6$, Br, $\ldots$, form a series of quasi-1D conductors characterized by the same crystallographic structure as  the Bechgaard salts  (TMTSF)$_2X$ series\cite{Delhaes79}. The difference between the two series lies in the chemical composition of  the TMTTF organic molecule  for which  the sulfur atoms are substituted in place of  the selenium in TMTSF. In consequence the amplitude of the dimerization  of TMTTF stacks   turns out to be  more pronounced  in the solid state,  along with   band parameters that are typically smaller than those  found in  (TMTSF)$_2X$ (see Sec.~\ref{Model})\cite{Grant83,Ducasse86}.  In normal pressure conditions, the (TMTTF)$_2X$ are thus more one-dimensional in character and also more correlated than (TMTSF)$_2X$  through essentially a stronger influence of electronic umklapp processes. 

 This is exemplified by an upturn in electrical resistivity at the intermediate temperature, $T_\rho$,  indicative of strong umklapp scattering that  evolves   towards an insulating 1D Mott  behaviour \cite{Coulon82,Laversanne84,Emery82,Giamarchi97}. Long-range ordered states can be found at much  lower temperature which can involve charge, spin and even lattice degrees of freedom\cite{Bourbon08}. A remarkable property of the series emerges   when sufficiently high pressure is applied to (TMTTF)$_2X$  which ultimately maps their physical properties   to that of (TMTSF)$_2X$ at low pressure.

\begin{figure}
\includegraphics[width=8cm]{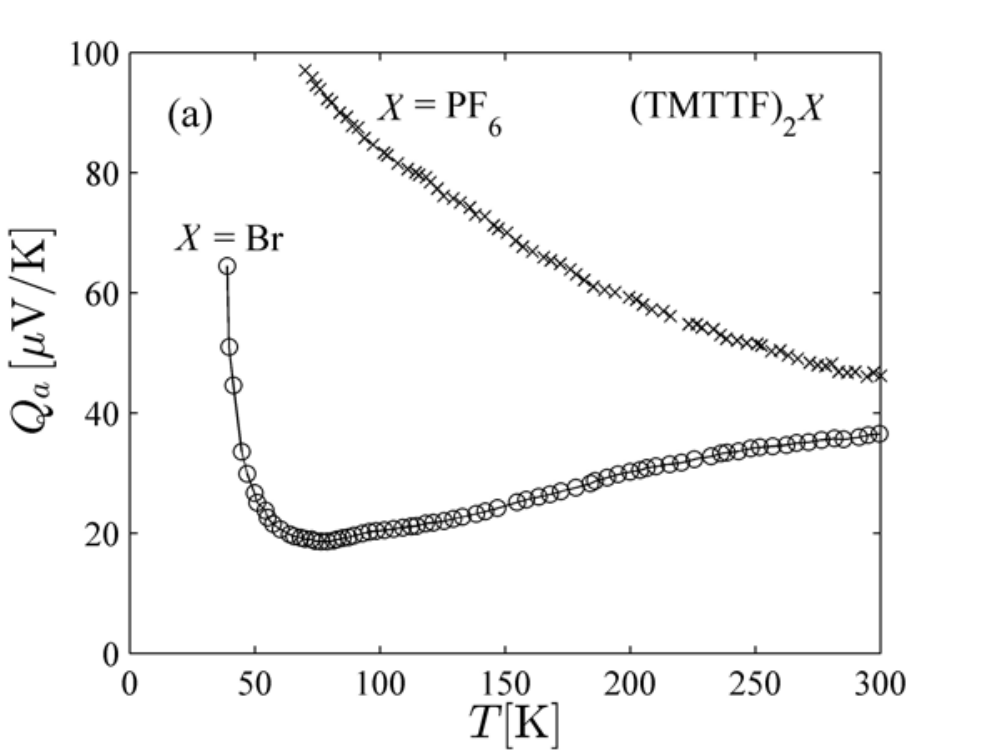}\\
\includegraphics[width=8cm]{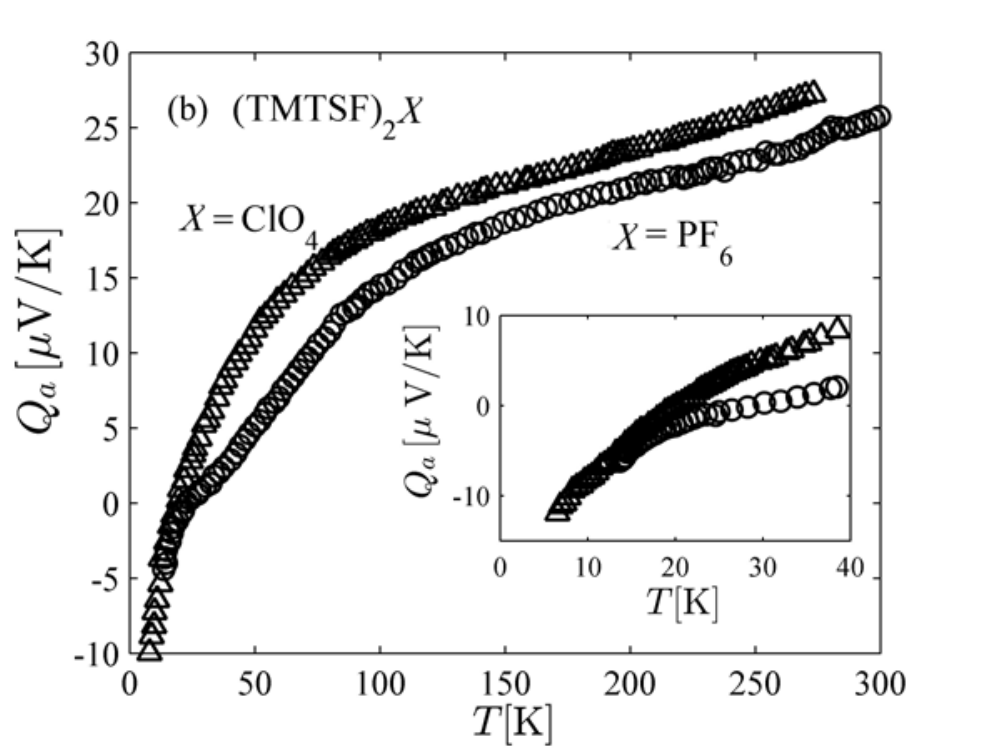}
 \caption{ Temperature dependence of Seebeck coefficients in (a): few (TMTTF)$_2X$ salts at ambient pressure (after Ref.\cite{Mortensen83}); (b):   (TMTSF)$_2X$ in the metallic state above $T_{\rm SDW}$, $X=$~PF$_6$, and $T_c$, $X=$~ClO$_4$. The inset shows the   sign reversal of the Seebeck coefficient in  low temperature domain    (after Ref.\cite{Chai09}).} 
\label{Qexp}
\end{figure}

 (TMTTF)$_2$PF$_6$ is a prototype compound of  the  Fabre series characterized by the temperature scale  $T_\rho \simeq 220$K,\cite{Coulon82,Laversanne84}. The measurements of    the Seebeck coefficient by Mortensen {\it et al.,}\cite{Mortensen83}  for this compound are reproduced in Fig.~\ref{Qexp}~(a). The data show a monotonic increase of the Seebeck coefficient with decreasing temperature. The increase is consistent with non metallic behaviour shown by resistivity in the same range of temperature\cite{Coulon82,Laversanne84}. 
 
The calculated results  for a compound like (TMTTF)$_2$PF$_6$  are displayed in Fig.~\ref{Qht} (b) when  in accordance with band calculations  \cite{Ducasse86}, smaller hopping terms   ($t/k_B= 1300$K, $t_\perp/k_B= 100$K, $E_F/k_B=1500$K, $t_\perp'/k_B =15$K)\footnote{We have chosen an effective range of values  for $t_\perp$ that is slightly lower than the one of band calculations,  in order to incorporate the  effect of the downward renormalization of $t_\perp$ that takes place at the approach of the Mott scale $T_\rho$. This effect is not taken into account by the present  RG calculations at the one-loop level\cite{Bourbon08,Sedeki12}} and larger amplitudes for  the bare umklapp ($g_3=0.15)$ are used.  With these figures, the instability at $T_{\rm SDW}$ occurs at much higher temperature ($T_{\rm SDW}=T_\rho\sim t_\perp$) and corresponds to the 1D Mott scale $T_\rho$ at the one-loop level of the RG\cite{Bourbon08,Emery82}.  The  important reduction of  the longitudinal hopping $t$ is responsible for a larger amplitude   of the Seebeck coefficient, which is mainly dominated by the band term $Q_a^0$ at high temperature, as shown by the dashed line of Fig.~\ref{Qht} (b). Note that this  term   surpasses the total $Q_a$  indicating  that the contribution of $Q_a^c$ coming from  collisions  is relatively small but negative at very high temperature. The resulting $Q_a$ then shows a smooth decrease with lowering temperature  contrary to observation in (TMTTF)$_2$PF$_6$. However, the effect  of the collision term   becomes quickly positive and to give rise to an upturn of the   Seebeck coefficient with lowering temperature as observed.  
 
 The case of (TMTTF)$_2$Br is also of interest since   along  the pressure axis of a generalized phase diagram including both families, this compound  is  chemically shifted  at about half distance between (TMTTF)$_2$PF$_6$ and the members of (TMTSF)$_2X$ series at low pressure\cite{Bourbon08}.  For  normal state properties for instance, this is illustrated by the   intermediate scale  $T_\rho\simeq 100$K seen in   resistivity    \cite{Coulon82,Klemme95}, in line  with a smaller dimerization of the organic stacks for  (TMTTF)$_2$Br. The temperature variation of the Seebeck coefficient   for the bromine salt is displayed in Fig.~\ref{Qexp}  \cite{Mortensen83}. At room temperature the  coefficient is smaller in amplitude compared to  (TMTTF)$_2$PF$_6$; it drops as temperature is lowered,   consistently with  the more pronounced metallic character of this salt in this temperature range. However, the variation is not linear in temperature but reveals an enhancement with respect to the free carrier situation.   Comes a minimum for temperature under  $T_\rho$,  followed by a rise that evolves  toward a characteristic  $1/T$ behaviour for $Q_a$ at sufficiently low temperature\cite{Mortensen83},  in congruence with a well defined insulating (Mott) gap.
 
 By using intermediate figures for the band parameters ($t= 2000$K, $t_\perp = 100$K, $E_F= 2200$K, $t_\perp'/k_B =15$K) and bare umklapp amplitude ($g_3= 0.08$), the amplitude of the calculated Seebeck coefficient at ambient temperature in Fig.~\ref{Qht} is   intermediate   between   (TMTTF)$_2$PF$_6$ and (TMTSF)$_2X$,  as shown in Fig.\ref{Qexp} (a)-(b). The calculated decrease  of $Q_a$, though enhanced compared to $Q_a^0$ due to inelastic scattering, is less rapid  than  observed. The flow to strong umklapp scattering then results in the upturn  in the Seebeck coefficient.

\subsection{The Bechgaard salts (TMTSF)$_2X$}
The Seebeck coefficient  measured for  the $X$=PF$_6$ and ClO$_4$ members of  the Bechgaard  is shown in Fig.~\ref{Qexp}~(b). Let us remind that in contrast with the sulfur based  (TMTTF)$_2X$ compounds at low pressure, $T_\rho$ is an irrelevant 1D scale for (TMTSF)$_2X$, since these materials  are metallic down to the temperature of onset of long-range order ($T_{{\rm SDW},c}\ll t_\perp$). In this range of temperature the system  becomes   effectively 2D  regarding one-particle coherence, though strongly anisotropic.
In  Fig.~\ref{Qexp}~(b) we reproduced the temperature dependence of    $Q_a$    obtained by Chai {\it et al.,}\cite{Chai09} (see also  Choi {\it et al.,}\cite{Choi02} Sun {\it et al.,}\cite{Sun08},  Gubser {\it  et al.}\cite{Gubser82} and Chaikin\cite{Chaikin83} ). The    (TMTSF)$_2$PF$_6$  shows  SDW ordering  at $T_{\rm SDW} \simeq 12.5$K, whereas for  (TMTSF)$_2$ClO$_4$   the anion (ClO$_4$) ordering in slow cooling conditions   pushes the compound on the  SC part of the phase diagram of series with $T_c\simeq 1.2$K\cite{Bechgaard81,Jerome82}.  

Close to ambient temperature, the Seebeck coefficient for both compounds reveals values relatively close to the calculated    band limit $Q_a^0$ given in  Fig.~\ref{Qht} using the (TMTSF)$_2X$  band parameters of Sec.~\ref{Model}. At lower temperature a positive enhancement with respect to a $T-$linear descent is observed for both compounds, in qualitative agreement with the one found in the present calculations in Fig.~\ref{Qht}. However, in contrast to predictions, no maximum for $Q_a$ is found experimentally; the data of Fig.~\ref{Qexp}~(b)   rather show  a shoulder like structure that precedes  the low temperature descent of the Seebeck coefficient. This suggests that the energy variation of the collision term is less rapid than predicted in this temperature range. It is not excluded, however, that such a maximum would show up if small pressure was applied to a compound like (TMTTF)$_2$Br [see Fig.~\ref{Qexp}~(a)], which would suppress $T_\rho$\cite{Klemme95} and shift the compound on the left hand side of (TMTSF)$_2X$ along the pressure axis\cite{Bourbon08}(see also the footnote in\footnote{It should be  stressed here that experiments are performed in constant pressure conditions, whereas the calculations are obtained at constant volume.  For relatively  soft materials like the organics, constant volume corrections to the data may modify the actual temperature dependence for a transport quantity like the Seebeck coefficient (See for example Ref.\cite{Moser98}).}).

For both compounds, the Fig.~\ref{Qexp}~(b) shows that the drop seen at  low temperature for the Seebeck coefficient  does not extrapolate toward zero, but  exhibits negative enhancement from $Q_a^c$ that evolves toward to a net  sign reversal of the Seebeck coefficient. The fact that the $T_c$ for (TMTSF)$_2$ClO$_4$ is small, this sign reversal can be followed down to the lowest temperature of the metallic state, as shown in the insert of Fig.~\ref{Qexp}~(b).   The sign reversal of the Seebeck coefficient occurs in the metallic state in the absence of a Fermi surface reconstruction that would transform the nature of carriers from  hole to electron type. Therefore the present calculations provide an avenue of explanation for this effect in  terms of an anomalous energy dependence of the inelastic umklapp scattering at the Fermi level that becomes electron like in character (see Fig.~\ref{dphidE}). This transformation takes its origin in the SDW fluctuations which act  as the source of enhancement of umklapp scattering. As we have seen in Sec.~\ref{RG}, the  development of these spin fluctuations can be  greatly enhanced over   sizable  intervals of temperature and  antinesting in the neighborhood of the quantum  critical point $t_\perp'^*$ [see for example Fig.~\ref{g3}~(b)]\cite{Bourbon09}, in qualitative correspondance with the spreading of the sign reversal of the  Seebeck coefficient found  in Figs.~\ref{Q} and \ref{Qint}. It is worth stressing that  NMR experiments for the temperature variation of the nuclear spin lattice relaxation rate have brought  considerable evidence for  the presence of spin fluctuations  for both compounds in the same temperature range and their   amplitude with pressure\cite{Creuzet87b,Bourbon84,Wzietek93,Brown15,Brown08,Shinagawa07}.
\section{Summary and concluding remarks}
In the work developed above, a derivation of the Seebeck coefficient in quasi-1D interacting electron systems has been carried out from a numerical solution of the linearized Boltzmann equation using  the renormalization group method for the evaluation of the electron-electron scattering matrix element. From  a  parametrization of the electron gas model compatible with the spin-density-wave to superconducting sequence of orderings found  in organic superconductors under pressure,  the temperature variation of the Seebeck coefficient in the metallic phase was  calculated.  It was shown to develop  marked deviations with respect to the hole band linear-$T$ prediction. These deviations found their striking expression in the quantum critical region of the metallic phase linked to  the juncture of antiferromagnetic and superconducting  orders. It is where the enhancement of the Seebeck coefficient undergoes a sign reversal, attributable to an anomalous low energy variation of the anisotropic  electron-electron scattering   time   becoming electron-like in character over most of the Fermi surface. Spin fluctuations, which act as a source of inelastic umklapp scattering, appear as a key determinant   for this sign reversal which occurs  in the absence of a Fermi surface reconstruction. It is only when the antinesting parameter, which simulates the role of  pressure in real materials like the organics, is tuned sufficiently far away from the quantum critical  point that a Fermi liquid type of enhancement of the Seebeck coefficient is recovered. 

The results were shown to capture many features of existing data in quasi-1D conductors like the Bechgaard salts (TMTSF)$_2X$, in particular the crossover to negative  values of  the   Seebeck coefficient at low temperature in the neighbourhood  of  their quantum critical point along the pressure axis. The size of the enhancement with respect to the band prediction is also fairly well taken into account, suggesting that electron-electron scattering in the presence of electron-hole asymmetry   mainly due to nesting alterations is likely to be the most important processes governing the temperature dependence of the thermoelectric response of these materials. The results of the calculations being obtained at arbitrary antinesting distance from the quantum critical point of the phase diagram,    can serve as a stimulus for   future  experiments of the Seebeck coefficient  in Bechgaard salts as function of applied hydrostatic pressure. Such a systematic study  is lacking so far. Following the example of previous electrical transport\cite{DoironLeyraud09,DoironLeyraud10} and NMR studies\cite{Creuzet87b,Brown08,Brown15,Creuzet85a}, it would be worthwhile to check if the Fermi liquid behaviour for the  Seebeck coefficient is recovered  sufficiently  far above the quantum critical pressure. 

The  present theory of the Seebeck coefficient was also applied to the more correlated Fabre (TMTTF)$_2X$ series, whose members   with centro-symetrical anions $X$ are known to become Mott insulators at relatively high temperature   in normal pressure conditions. Stronger umklapp scattering  and narrower bands  characterize these materials. This precipitates  an instability toward an insulating state  at much higher temperature    and accordingly, yields a pronounced  enhancement of   the Seebeck coefficient that is present in experiments.

 \acknowledgments
 The authors would like to thank H. Bakrim and A. Sedeki  for their valuable comments on computing aspects of this work.
 C. B. thanks the National Science and Engineering Research Council  of Canada (NSERC) and the R\'eseau Qu\'eb\'ecois des Mat\'eriaux de Pointe (RQMP) for financial support. Computational resources were provided
by the R\'eseau qu\'eb\'ecois de calcul de haute performance
(RQCHP) and Compute Canada.

\appendix


\section{Linearized Boltzmann equation}
In the presence of an external longitudinal thermal gradient  and  induced electric field, the linearized Boltzmann equation for the normalized deviations $\bar{\phi}_{\bm{k}}$ ($= \bar{\phi}^{j={\cal E},T}_{\bm{k}}$) can be put in  the single  form of Eq.~(\ref{BEQ}):
\begin{align}
\label{LBE_a}
{{\cal L}}\bar{\phi}_{\bm{k}} = \sum_{i=1}^4 \sum_{\bm{k'}}{\cal L}^{[i]}_{\bm{k},\bm{k'}} \bar{\phi}_{\bm{k'}}=1,\end{align}
The collision operator is expressed as the sum of four terms, 
\begin{widetext}
\begin{align}
\label{L_a}
\sum\limits^4_{i=1} \mathcal{L}^{[i]}_{\bm{k},\bm{k^\prime}} = \dfrac{1}{(L N_P)^2} &\sum\limits_{\bm{k}_2,\bm{k}_3,\bm{k}_4}  {1\over 2}|\langle\bm{k},\bm{k}_2\vert g_3 \vert\bm{k}_3,\bm{k}_4\rangle   
 -\langle\bm{k},\bm{k}_2\vert g_3 \vert\bm{k}_4,\bm{k}_3\rangle|^2 
\times \frac{2\pi}{\hbar}  \delta_{\bm{k}+\bm{k}_2,\bm{k}_3+\bm{k}_4+ p \bm{G}}  \nonumber \\
&\times  \delta(\varepsilon^p_{\bm{k}}+\varepsilon^{p_2}_{\bm{k}_2}-\varepsilon^{p_3}_{\bm{k}_3}-\varepsilon^{p_4}_{\bm{k}_4}) \dfrac{f^0(\bm{k}_2)[1-f^0(\bm{k}_3)][1-f^0(\bm{k}_4)]}{[1-f^0(\bm{k})]} (\delta_{\bm{k},\bm{k^\prime}} + \delta_{\bm{k}_2,\bm{k^\prime}} - \delta_{\bm{k}_3,\bm{k^\prime}} - \delta_{\bm{k}_4,\bm{k^\prime}}).
\end{align}
\end{widetext} 
The amplitude of umklapp vertex are evaluated by the RG  in the framework of  the quasi-1D electron gas model, 
\begin{align}
\label{}
   \langle\bm{k}_1,\bm{k}_2\vert g_3 \vert\bm{k}_3,\bm{k}_4\rangle = & \,\pi\hbar v_F g_3(\boldsymbol{k}_{F,1}^p,\boldsymbol{k}_{F,2}^p;\boldsymbol{k}_{F,3}^{-p},\boldsymbol{k}_{F,4}^{-p})\cr
= & \, \pi\hbar v_F g_3(k_{\perp1},k_{\perp2};k_{\perp3},k_{\perp4}),
\end{align}
where the position on the Fermi surface  is parametrized by the transverse wavevectors.
To solve the equation with the explicit form of matrix elements shown in equation \ref{L_a}, we separate the momentum conservation constraint into longitudinal and transverse components, \cite{Gorkov98}
\begin{align}
& \delta_{\bm{k}+\bm{k}_2,\bm{k}_3+\bm{k}_4+ p \bm{G}} =  \  \delta_{k_\perp + k_{\perp2},k_{\perp3}+k_{\perp4}} \cr
&  \ \ \ \ \times  \frac{2\pi}{L} \hbar v_F  \delta( \varepsilon^p_{\bm{k}}+\varepsilon^{p_2}_{\bm{k}_2}+\varepsilon^{p_3}_{\bm{k}_3}+\varepsilon^{p_4}_{\bm{k}_4} - \Sigma),\cr
\end{align}
where $\Sigma = \epsilon_\perp(k_\perp)+\epsilon_\perp(k_{\perp2})+\epsilon_\perp(k_{\perp3})+\epsilon_\perp(k_{\perp4})$. The summation over the momentum vectors can be written as
\begin{eqnarray}
\label{sum}
\dfrac{1}{L N_P}\sum\limits_{\bm{k}} =  \sum_p \int \dfrac{d\varepsilon^p_{\bm{k}}}{2\pi \hbar v_F} \frac{1}{N_P} \sum\limits_{k_\perp}.
\end{eqnarray}
Carrying out the integration over $\varepsilon^{p'}_{\bm{k^\prime}}$ and by rearranging the terms, we arrive  at the final equation,
\begin{widetext}
\begin{eqnarray}
\label{LBEb}
 {{\pi}\over \beta\hbar}   {1\over N_P^2}\!  &\sum\limits_{k^\prime_{\perp},k_{\perp3}, k_{\perp4}}& \Big\{ {\vert {g_3}(k_\perp , k_{\perp3}+k_{\perp4}-k_\perp; k_{\perp3},k_{\perp4}) - {g_3}(k_\perp , k_{\perp3}+k_{\perp4}-k_\perp; k_{\perp4},k_{\perp3}) \vert}^2  \delta_{k_\perp,k^\prime_\perp}  \nonumber \\
&\times & \dfrac{\beta \Sigma^\prime /4 \cosh(\beta E/2)}{\cosh(\beta(\Sigma^\prime /4 - E/2)) \sinh(\beta \Sigma^\prime /4)} + {\vert  {g_3}(k_\perp , k^\prime_\perp; k_{\perp3},k_{\perp4}) -  {g_3}(k_\perp , k^\prime_\perp; k_{\perp4}, k_{\perp3}) \vert}^2 \nonumber \\
&\times & \dfrac{\beta \Sigma^{\prime\prime} /4 \cosh(\beta E/2)}{\cosh(\beta(\Sigma^{\prime\prime} /4 - E/2)) \sinh(\beta \Sigma^{\prime\prime} /4)} \delta_{k_\perp + k_{\perp}^\prime,k_{\perp3}+k_{\perp4}} - 2 {\vert  {g_3}(k_\perp, k_{\perp3}; k_{\perp}^\prime, k_{\perp4}) - {g_3}(k_\perp , k_{\perp3}; k_{\perp4}, k_{\perp}^\prime) \vert}^2  \nonumber \\
&\times & \dfrac{\beta \Sigma^{\prime\prime} /4 \cosh(\beta E/2)}{\cosh(\beta(\Sigma^{\prime\prime} /4 - E/2)) \sinh(\beta \Sigma^{\prime\prime} /4)} \delta_{k_\perp + k_{\perp3}, k^\prime_\perp + k_{\perp4}} \Big\}  \bar{\phi}_{E , k^\prime_\perp} = 1, 
\end{eqnarray}
\end{widetext}
where {$\Sigma' = \epsilon_\perp(k_\perp) + \epsilon_\perp(k_{\perp3}+k_{\perp4}-k_\perp) + \epsilon_\perp(k_{\perp3}) + \epsilon_\perp(k_{\perp4})$} and {$
\Sigma^{\prime\prime} = \epsilon_\perp(k_\perp) + \epsilon_\perp(k_\perp^\prime) + \epsilon_\perp(k_{\perp3}) + \epsilon_\perp(k_{\perp4})$}. By inserting the RG results of Sec. \ref{RG} for the momentum resolved umklapp scattering, the numerical solution of (\ref{LBEb})  for $\bar{\phi}_{E,k_\perp}$ allows the evaluation of the scattering contribution $Q_a^c$ to the Seebeck coefficient (\ref{thermopower}).

\bibliography{/Users/cbourbon/Dossiers/articles/Bibliographie/articlesII.bib}

\begin{thebibliography}{65}
\bibitem [{Not()}]{NoteSeebeck1}%
  \BibitemOpen
  \href@noop {} {}\bibinfo {note} {{This much longer time scale for scattering
  processes would be rather compatible with the so-called zero frequency mode
  seen in optical conductivity of the Bechgaard salts in their low temperature
  metallic state \cite{Dressel96}}}\BibitemShut {NoStop}%
\end{thebibliography}

\end{document}